\documentclass[letterpaper, 10 pt, conference, twocolumn, titlepage]{ieeeconf}
\IEEEoverridecommandlockouts                              % This command is only
\usepackage{amsmath} % assumes amsmath package installed
\usepackage{amssymb}  % assumes amsmath package installed
\usepackage{algorithm,algorithmic}
\usepackage{graphicx}
\usepackage{verbatim} % used to define comment
\usepackage{psfrag}
\usepackage{subfigure}
\usepackage{caption}

\newtheorem{theorem}{Theorem}
\newtheorem{lemma}[theorem]{Lemma}
\newtheorem{definition}{Definition}
\newtheorem{example}{Example}

\def\supertiny{\font\supertinyfont=cmr10 at 3pt \relax \supertinyfont}
\newcommand{\mathscriptsize}[1]{\mbox{\tiny $#1$}} 
\newcommand{\mathsupertiny}[1]{\mbox{\supertiny $#1$}}

\begin{document}

\title{\LARGE \bf
Low-Complexity Non-Uniform Demand Multicast Network Coding Problems
}

\author{Joseph~C.~Koo and John~T.~Gill,~III% <-this % stops a space
\thanks{J. Koo and J. Gill are both with the Department of Electrical Engineering, 
        Stanford University, Stanford, CA  94305, USA.
        E-mail: {\tt\small \{jckoo, gill\}@stanford.edu}}%
}

\maketitle
\thispagestyle{empty}
\pagestyle{empty}

%%%%%%%%%%%%%%%%%%%%%%%%%%%%%%%%%%%%%%%%%%%%%%%%%%%%%%%%%%%%%%%%%%%%%%%%%%%%%%%%
\begin{abstract}

The non-uniform demand network coding problem is posed as a
single-source and multiple-sink network transmission problem where the
sinks may have heterogeneous demands.  In contrast with multicast
problems, non-uniform demand problems are concerned with the amounts
of data received by each sink, rather than the specifics of the
received data.  In this work, we enumerate non-uniform network demand
scenarios under which network coding solutions can be found in
polynomial time.  This is accomplished by relating the demand problem
with the graph coloring problem, and then applying results from the
strong perfect graph theorem to identify coloring problems which can
be solved in polynomial time.  This characterization of
efficiently-solvable non-uniform demand problems is an important step
in understanding such problems, as it allows us to better understand
situations under which the NP-complete problem might be tractable.

\end{abstract}

%%%%%%%%%%%%%%%%%%%%%%%%%%%%%%%%%%%%%%%%%%%%%%%%%%%%%%%%%%%%%%%%%%%%%%%%%%%%%%%%
\section{Introduction}
\label{sec-intro}

\begin{figure}[htbp]
\centering
\subfigure[Butterfly network.]{
\psfrag{s}[c][c]{\footnotesize $s$}
\psfrag{u}[c][c]{\footnotesize $u$}
\psfrag{v}[c][c]{\footnotesize $v$}
\psfrag{w}[c][c]{\footnotesize $w$}
\psfrag{x}[c][c]{\footnotesize $x$}
\psfrag{t1}[c][c]{\footnotesize $t_{\mathscriptsize{1}}$}
\psfrag{t2}[c][c]{\footnotesize $t_{\mathscriptsize{2}}$}
\psfrag{b1}[c][c]{\tiny $b_{\mathsupertiny{1}}$}
\psfrag{b2}[c][c]{\tiny $b_{\mathsupertiny{2}}$}
\psfrag{b1xxxb2}[c][c]{\tiny $b_{\mathsupertiny{1}} \oplus b_{\mathsupertiny{2}}$}
\includegraphics[height=1.5in]{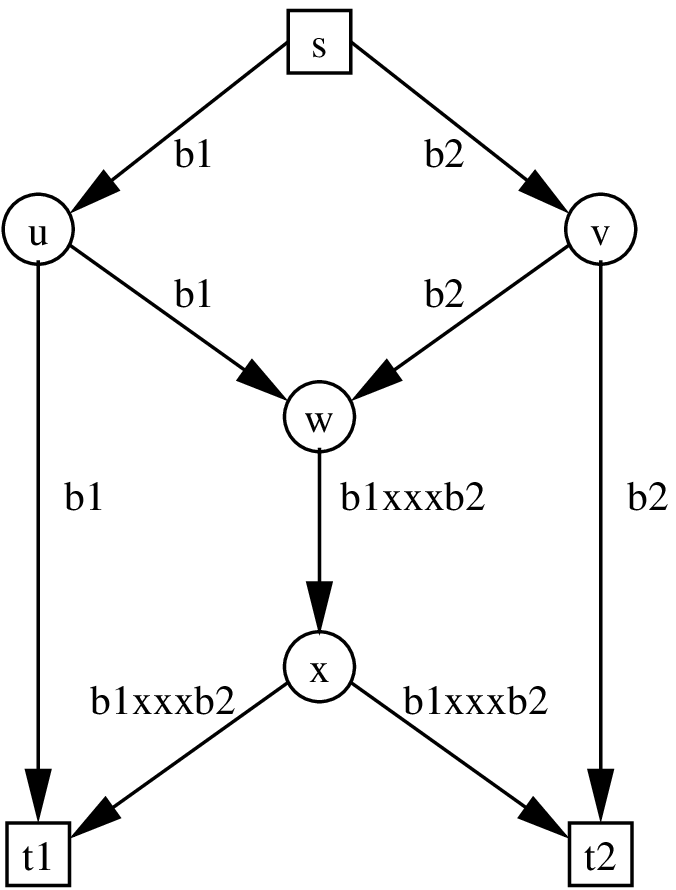}
\label{fig:butterfly_network}
}
\hspace{0.1in}
\subfigure[Extended butterfly network.]{
\psfrag{s}[c][c]{\footnotesize $s$}
\psfrag{u}[c][c]{\footnotesize $u$}
\psfrag{v}[c][c]{\footnotesize $v$}
\psfrag{w}[c][c]{\footnotesize $w$}
\psfrag{x}[c][c]{\footnotesize $x$}
\psfrag{y}[c][c]{\footnotesize $y$}
\psfrag{z}[c][c]{\footnotesize $z$}
\psfrag{t1}[c][c]{\footnotesize $t_{\mathscriptsize{1}}$}
\psfrag{t2}[c][c]{\footnotesize $t_{\mathscriptsize{2}}$}
\psfrag{b1}[c][c]{\tiny $b_{\mathsupertiny{1}}$}
\psfrag{b2}[c][c]{\tiny $b_{\mathsupertiny{2}}$}
\psfrag{b3}[c][c]{\tiny $b_{\mathsupertiny{3}}$}
\psfrag{b1xxxb2}[c][c]{\tiny $b_{\mathsupertiny{1}} \oplus b_{\mathsupertiny{2}}$}
\includegraphics[height=1.5in]{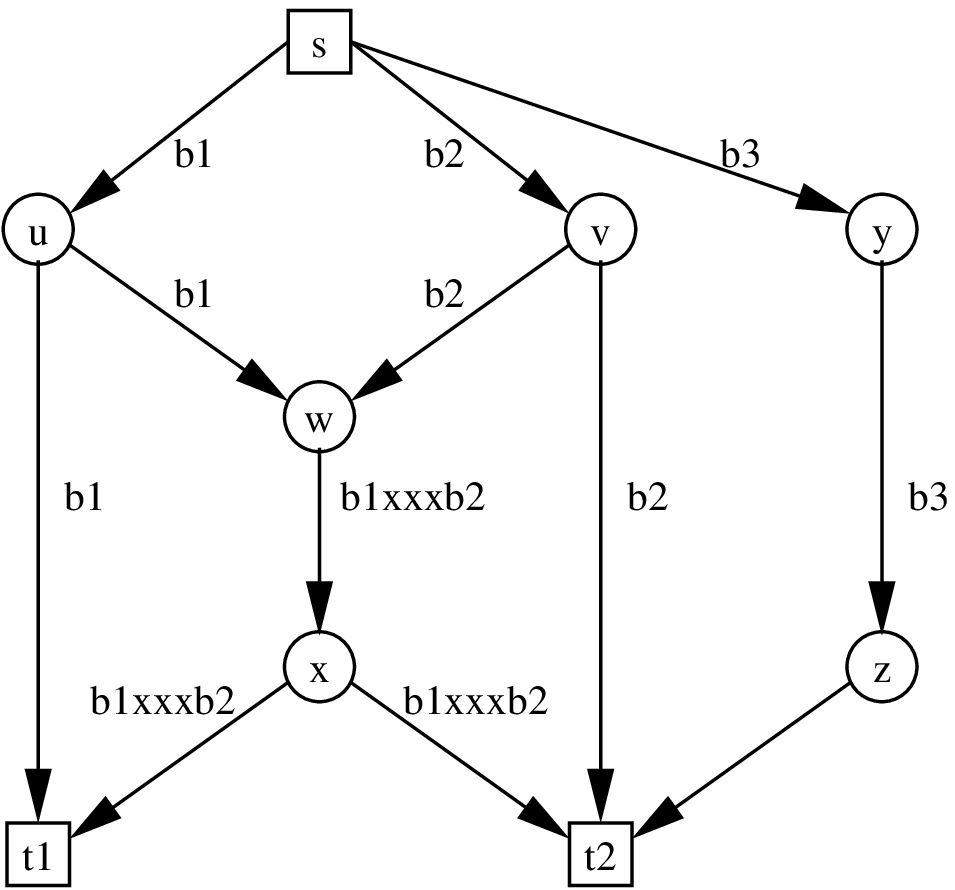}
\label{fig:extended_butterfly_network}
}
\renewcommand{\captionfont}{\footnotesize}
\caption{The butterfly network (Figure~\ref{fig:butterfly_network},
adapted from~\cite{ahlswede:network_info_flow}) shows how network
coding increases possible data rates.  Source~$s$ is transmitting data
streams~$b_1$ and~$b_2$ to both sinks~$t_1$ and~$t_2$.  The network
code has node~$w$ sending $b_1 \oplus b_2$ on its outgoing link.  As a
result, $s$ can transmit both streams to both sinks in one time
period, instead of the two time periods required by routing solutions.
The extended butterfly network
(Figure~\ref{fig:extended_butterfly_network}) has an additional path
to sink~$t_2$ which can be utilized for sending additional data.
Using network coding and taking advantage of the extra path through
nodes~$y$ and~$z$, source~$s$ can transmit three streams ($b_1$,
$b_2$, and~$b_3$) to sink~$t_2$ simultaneously with two streams ($b_1$
and~$b_2$) to sink~$t_1$.}
\label{fig:butterfly_network_both}
\end{figure}

Network coding has been shown to enable higher transmission rates
across communication networks, when compared against routing.  This is
because network coding allows data flows toward different sinks to
share the same links, and---through appropriate coding of
symbols---have the sinks still be able to decode out these
disparate flows.  In the butterfly network example first
proposed by Ahlswede \textit{et al.}~\cite{ahlswede:network_info_flow}
(see Figure~\ref{fig:butterfly_network}), if the input data at
node~$w$ are coded together, it is possible to multicast two
streams of information~$b_1$ and~$b_2$ from the source~$s$ to both
sinks~$t_1$ and~$t_2$ within a single time period.  The benefits of
allowing coding at nodes are evident; under routing, multiple
time periods would be required to send both streams to
both sinks.  The authors show that in any network with a single source
and multiple sinks, the information rate can achieve the
minimum (over all sinks) of the maximum flow to the sink nodes.  In
subsequent work, Li \textit{et al.}~\cite{li:linear_net_coding} prove
that linear network codes are sufficient for multicast, and Jaggi
\textit{et al.}~\cite{jaggi:polytime_algs_multicast_code_constr} give
a polynomial-time algorithm for constructing such linear codes.

Following the quick successes of characterizing and developing
algorithms for multicast network coding problems, there has been
much work concerning the construction of network codes for more
general scenarios---although this has proven to be much more
difficult.  Koetter and
M\'{e}dard~\cite{koetter:algebraic_approach_net_coding} give an
algebraic characterization for achievable linear network codes, but
prove that checking for the existence of such codes requires running
time which is not polynomially bounded.  Then, Rasala Lehman and
Lehman~\cite{rasalalehman:complex_class_net_info_flow} prove that for
most network coding scenarios, finding linear network codes to satisfy
arbitrary source and demand requirements is NP-hard.  Of relevance to
the current work is the problem of constructing network codes to send
data from a single source to multiple sinks with arbitrary demands
(potentially with different demands by different sinks).

In this work, we study networks where the single source may send data
to multiple sinks at unequal rates.  The motivation for this can be
seen in the extended butterfly network of
Figure~\ref{fig:extended_butterfly_network}.  Here, the traditional
butterfly network is augmented with an additional path between the
source~$s$ and sink~$t_2$.  Within a single time period, at most two
streams can be transmitted to sink~$t_1$, but it is possible to
transmit more than two streams to sink~$t_2$.  If sink~$t_2$ is
constrained to only receive two streams, then available capacity is
wasted.

The problem of sending unequal-rate data from a single source to
multiple sinks has two flavors: the multiple multicast connections
problem~\cite{lun:net_coding_cost_criterion} and the non-uniform
demand problem~\cite{cassuto:non_uniform_demands}.  In the multiple
multicast connections problem, the sinks are allowed to receive data
at different rates, but the subset of information demanded by each
particular sink---while arbitrary---is identified in advance.  On the
other hand, in the non-uniform demand problem, the amount of
information a particular sink must be able to receive is specified in
advance, but it does not matter which specific pieces of information
are received.  This is a scenario where the source has a large set of
messages which it wishes to send to the sinks, and each particular
sink wishes to receive a subset of the source's messages; however, the
requirement at each sink is only that the messages it receives is a
subset of a particular size rather than a requirement of receiving
some specific subset of messages.  This problem can be understood as a
relaxed version of the multiple multicast connections network coding
problem, since if it is known that a sink is unable to receive a
subset of size $n$, then any specific demand for a subset of size $n$
can automatically be rejected as impossible.  Conversely, if it is
known that a sink is able to receive some subset of size $n$, then it
may be possible to find a network coding solution with a specific
demand for that sink which is of size $n$.  The non-uniform demand
problem makes it possible to determine the maximum possible data rates
that can be received by the sinks.

The non-uniform demand problem was originally investigated by Cassuto
and Bruck~\cite{cassuto:non_uniform_demands}.  For demand scenarios
where all sinks require the same high rate except for two sinks
demanding some lower rate, the authors prove that it is possible to
satisfy these demands in all cases---and linear codes are sufficient.
They also give limited conditions for the achievability of non-uniform
demands when there are more than two lower demanded rate sinks along
with any number of higher rate equal-rate sinks.  In our work, we more
specifically consider the case of non-uniform demands where each
[possibly heterogeneous] sink demands a data rate of its own
maximum-flow (i.e., maximum point-to-point rate from the source).
From this seemingly more restrictive setting, however, we are able to
describe a larger class of networks for which the non-uniform demand
problem is solvable (and solvable in polynomial time), thus enabling
non-uniform demand network coding to be more widely applicable.

Although the class of network demand scenarios for which we give
polynomial-time solutions is not exhaustive, it may be difficult to
enumerate the conditions more generally.  This is because the
non-uniform demand network coding problem is
NP-hard~\cite{cassuto:non_uniform_demands}.  We give an alternate
proof of the NP-hardness of the non-uniform demand network coding
problem (see Appendix), which proves this result for slightly
different demand scenarios than those addressed
in~\cite{cassuto:non_uniform_demands}.

\subsection{Related Work}
\label{subsec-relatedwork}

A technique which we use is that of transmitting data along paths, or
through flows.  This approach has been widely used in the network
coding literature, and has enabled many significant results.  In Jaggi
\textit{et al.}~\cite{jaggi:polytime_algs_multicast_code_constr}, the
polynomial-time algorithms for multicast problems rely on the concept
of sending data down [perhaps overlapping] paths.
In~\cite{fragouli:info_flow_decomp}, Fragouli and Soljanin give a
decomposition of networks into flows, in order to model data
transmission in a network more simply.  Using this decomposition and a
graph coloring formulation, alphabet size bounds for any network code
are then proven.  Although the flow-based and path-based approaches
are similar in many ways, the two techniques differ in that the
flow-based approach creates a new flow every time a piece of data is
transformed by coding, whereas the path-based approach keeps track of
each piece of data as it is sent individually down a path, even if any
transformations get applied to the data.  We shall use a path-based
approach.

We briefly mention some results regarding the multiple multicast
connections problem, since achievable solutions for such problems are also
achievable for the non-uniform demand problem with the same demanded rates.
(Of course, the reverse is not always true.) Many of these results consider the
case of two sinks.  In~\cite{ngai:multisource_net_coding_two_sinks}, after
enumeration of all possible scenarios, the authors conclude that in the case of
two sinks with differing rates, linear coding is sufficient.  The same
conclusion is made
in~\cite{ramamoorthy:single_source_two_terminal_network_net_coding}, although
the authors use a different approach which considers a path-based enumeration.
A characterization of the achievable data rate region using network coding is
given for the two sink case.  For more than two sinks, conditions under which
solutions exist for the multiple connections problem have not been enumerated.

Separately from the multiple multicast connections problem, the
non-uniform demand problem itself has also been the subject of study.
In~\cite{cassuto:non_uniform_demands}, Cassuto and Bruck introduce the
problem and give some results concerning the achievability of the
individual max-flow rates to each sink.  In our work, we address
similar guarantees but for a wider class of network demands.  The
authors in~\cite{cassuto:non_uniform_demands} also prove that the
non-uniform demand problem is NP-hard, using a reduction from a 3-CNF
problem, although the demand problems for which their result holds
have network coding solutions that do not fully utilize the available
data rates (in our terminology, the solutions are not saturating).  We
supplement their proof by considering whether or not networks in which
the network coding solutions use all possible paths are still
difficult to solve.  We take a similar approach by considering
contamination amongst data; however, we do not allow intermediate
decoding as they do.  The non-uniform demand problem is also studied
in~\cite{chekuri:achiev_rates_non_uniform_demands}, which considers
the case where demands are allowed to be relaxed in the solution.  For
general networks, the authors give bounds on the fraction of max-flow
rate which is achievable, and show networks for which the bounds are
tight.  We take a different approach and instead characterize specific
network demand scenarios for which the max-flow rate can be achieved.

\subsection{Outline of Paper}
\label{outline}

In this paper, we will investigate the case of non-uniform demand
network coding in which each sink receives data at its individual
point-to-point (i.e., max-flow min-cut) capacity rate.  In
Section~\ref{sec-notationdef}, we discuss useful notation.  In
Section~\ref{sec-nonunifdemandassign}, we define our approach to the
problem, and analyze some of its characteristics.  Following that, in
Section~\ref{sec-streamassignalg}, we give an algorithm for
determining if a non-uniform demand solution exists, and discuss some
of its performance issues.  Using this algorithm, we characterize in
Section~\ref{sec-effnonunifdemand} a class of networks for which the
non-uniform demand network coding problem can be solved in polynomial
time.  Of course, not all non-uniform demand network coding problems
can be solved efficiently; in the Appendix, we give an alternate proof
of the NP-completeness of the non-uniform demand problem which
accounts for the demand scenarios we are considering.

%%%%%%%%%%%%%%%%%%%%%%%%%%%%%%%%%%%%%%%%%%%%%%%%%%%%%%%%%%%%%%%%%%%%%%%%%%%%%%%%
\section{Notation and Definitions}
\label{sec-notationdef}

We will consider a directed acyclic network graph $G = (V,E)$.  Each
sink will be indexed as $j \in \{1,2,\ldots,t\}$, where~$t$ is the
total number of sink nodes.  (Recall that there is only a single
source node~$s$.) Because the graph is acyclic, there exists a partial
ordering of the nodes starting from the source~$s$.  A partial
ordering of the edges can be constructed based on the ordering of the
nodes from which the edges originate; for edges $e = (v,w)$ and $e' =
(v',w')$, we denote $e \preceq e'$ if and only if $v \preceq v'$ in
the partial ordering of nodes.

For a particular sink~$j$, we define~$\mathcal{P}_j$ as the set of
paths associated with sink~$j$.  These are unit-capacity edge-disjoint
paths from the source~$s$ to sink~$j$ and can be determined from
maximum-flow algorithms such as the Ford-Fulkerson augmenting path
algorithm~\cite{ford:maxflow}.  Paths $p \in \mathcal{P}_j$ are given
as the elements of $\mathcal{P}_j = \{p_{j,1}, p_{j,2}, \ldots,
p_{j,n_j}\}$, where $n_j = |\mathcal{P}_j|$ is the number of paths to
sink~$j$.  Call the set of all paths $\mathcal{P} = \bigcup_{j=1}^{t}
\mathcal{P}_j$.  In contrast to much of the literature on network
coding for multicast, we will consider the \emph{maximum} of the
max-flows (instead of the minimum of the max-flows) and denote this
quantity~$n$, so $n = \max_j |\mathcal{P}_j| = \max_j n_j$.

In order to keep track of data that overlaps onto paths with other
sinks, we introduce the concepts of contamination and contaminating
set.  We say that \emph{contamination} from path~$p_{jk}$ onto
path~$p_{j'k'}$ occurs when data transmitted on~$p_{jk}$ gets combined
into the data which is supposed to be transmitted on~$p_{j'k'}$.  This
can occur, for example, when data on paths~$p_{jk}$ and~$p_{j'k'}$ are
coded together in order to be transmitted across an edge where the two
paths overlap.  Then the \emph{contamination set} of~$p_{jk}$ is the
set of all paths which experience contamination due to data
from~$p_{jk}$.  If we call~$\mathcal{D}_{jk}(e)$ as the set of paths
which are contaminated by path~$p_{jk}$ downstream of edge~$e$,
then~$\mathcal{D}_{jk}(e)$ can be defined recursively as follows:
\[
  \mathcal{D}_{jk}(e) = \bigcup_{p_{j'k'}} \left\{ p_{j'k'} \cup
  \left( \bigcup_{e' \succ e} \mathcal{D}_{j'k'}(e')\right) \right\}
\mbox{,}
\] 
where the union over~$p_{j'k'}$ is over all paths~$p_{j'k'}$ which
overlap path~$p_{jk}$ at edge~$e$ (and $j' \ne j$).  Then
$\mathcal{D}_{jk} = \bigcup_{e \in E} \mathcal{D}_{jk}(e)$ gives the
contamination set of~$p_{jk}$.  This definition accounts for a data
stream on a path to contaminate onto paths that it does not explicitly
overlap, due to contamination being spread from path to overlapping
path.

We also wish to keep track of the particular data streams sent to each
of the sinks.  A stream is defined as the identifier of the data that
is being transmitted down a particular path---as opposed to the
identifier of the path itself.  We identify a particular stream with
the index $i \in \{1,2,\ldots,n\}$, which means that a path is only
allowed to transmit a stream from the set $\{1,2,\ldots,n\}$.  Each
stream represents one information unit, of which only~$n$ unique
information units are allowed to be transmitted.  This restriction is
not prohibitive, since~$n$ is the maximum max-flow.  Each sink
receives a subset of the same~$n$ streams, so different sinks will
likely receive many of the same streams.  The number of distinct
streams a particular sink receives is its data rate, since each
unit-capacity edge-disjoint path can transmit at most only a single
data stream.

We also define decodable and saturating solutions.
\begin{definition}
A \emph{decodable solution} to a network coding problem is one in
which every sink is able to decode all of the information which is
intended to be sent to it.  In the example of streams assigned to
paths, a decodable solution is one in which every sink can recover all
of the streams which are assigned on paths to that sink.
\end{definition}
\begin{definition}
A \emph{saturating solution} is an assignment $f:\mathcal{P} \to
\{1,2,\ldots,n\}$ from paths $p_{jk} \in \mathcal{P}$ to streams
$\{1,2,\ldots,n\}$, such that for each $j \in \{1,2,\ldots,t\}$,
\[
  f(p_{jk}) \ne \emptyset, \ \forall k \in \{1,2,\ldots,n_j\}
\]
and
\[
  f(p_{jk}) \ne f(p_{jk'}), \ \forall k \ne k'
\mbox{.}
\]
\end{definition}
That is, a saturating stream assignment is a stream assignment in
which all paths to every sink are assigned some data stream; no path
is left unassigned.  Moreover, any streams assigned to different paths
to the same sink must be distinct.  Otherwise, if two paths carried
the same stream, one of the paths is redundant and does not carry
additional information. Thus, a saturating stream assignment is one in
which each sink~$j$ achieves its maximum possible data rate of~$n_j$.

We briefly mention the concept of intermediate decoding.
Specifically, for the network codes we are considering, we do not
allow intermediates nodes (i.e., nodes which are neither source nor
sink) to decode data and retransmit only a part of the data on its
outgoing links.  In other words, intermediate nodes are not allowed to
remove any contamination which it might receive on its incoming links,
even if it possesses enough information to decode out the
contamination.  Although this condition may prevent certain network
codes from being considered, it is still general enough that except
for certain cases, we should be able to find the appropriate network
coding solution if it exists.

\begin{definition}
The \emph{non-uniform demand problem} is the following solvability
problem (adapted from~\cite{cassuto:non_uniform_demands}):
Given a directed acyclic network graph $G = (V,E)$ (where each edge has
capacity~$1$), source~$s$, sinks $j \in \{1,2,\ldots,t\}$,
and demand function $d:\{1,2,\ldots,t\} \to \{1,2,\ldots,n\}$
(where~$d(j)$ is the demanded rate of sink~$j$), is there a network
coding solution such that for all~$j$, sink~$j$ receives information
at a rate~$d(j)$?
\end{definition}

%%%%%%%%%%%%%%%%%%%%%%%%%%%%%%%%%%%%%%%%%%%%%%%%%%%%%%%%%%%%%%%%%%%%%%%%%%%%%%%%
\section{The Non-Uniform Demand Stream Assignment Problem}
\label{sec-nonunifdemandassign}

The goal is to determine whether or not an assignment of data streams
to paths can give a decodable network solution.
\begin{definition}
The \emph{non-uniform demand stream assignment problem} is the
following:  Given a network graph $G = (V,E)$ and a decomposition into
paths, is there an assignment of streams to paths, such that no
intermediate decoding occurs, and the solution is both saturating and
decodable?
\end{definition}

We establish necessary and sufficient conditions for a network to have
a saturating and decodable solution.
\begin{theorem}
\label{thm:sat_decode_cond}
Given a set of paths between the source and the sinks, and if no
intermediate decoding is allowed, there exists a saturating and
decodable solution if and only if all streams which contaminate onto
paths to a particular sink have also been assigned to some other path
to the same sink.

\begin{proof}
Necessity follows from the fact that if the solution is decodable,
then each sink can separate out all streams and all contamination sent
to it.  For a particular sink, each path carries an assigned stream
mixed with the contamination from along that path.  Because no
intermediate decoding is allowed, all contamination arrives at the
sink, but arrives mixed in with the assigned streams on the respective
paths.  If there are~$n_j$ paths to the sink~$j$ and the solution is
saturating, then there are~$n_j$ unique streams assigned on paths to
the sink.  Now, if a contaminant is not also assigned to some other
path to that sink, then that means that there is data from at least
$n_j + 1$ streams  on inputs to the sink (the assigned~$n_j$ streams
of data plus at least one more data stream from the contamination).
However, because there are only~$n_j$ paths from the source to that
sink, where each path supports a data rate of~$1$, that means that no
more than~$n_j$ unique data streams can be received by the sink (or
else the max-flow condition would be violated).  Thus, any situation
where more than~$n_j$ data streams (perhaps mixed) can be seen by the
sink is a situation where fewer than~$n_j$ data streams can be decoded
successfully by the sink, and the solution is either not decodable or
not saturating.

For each sink, it is straightforward to show sufficiency of assigning
the contamination onto a path to that sink as the primary stream on
another of its paths, in order to guarantee decodability.  If no
intermediate decoding is allowed, all contamination arrives at the
sink.  If no other path has been assigned the same data stream, then
there is no way to determine (or even have partial knowledge) of the
data due to the contamination in order to either utilize or remove it.
Thus, there must be an assigned stream on some other path to that sink
which provides this information.
\end{proof}
\end{theorem} 

The concept of saturation is important, so that it is possible to
state (using max-flow theorems) that contamination without a
corresponding assigned stream can not be removed, as there will not be
enough flow to support this additional data.  Saturation is also
useful because it enables us to determine whether or not the maximum
data rate actually utilizes all paths, without repetitive data
streams.  In fact, if data is repeated (i.e., multiple paths to the
same sink are assigned the same stream), then one of the multiple
paths could be shut off with no harm to the data rate toward that
sink, and possibly even increasing data rates across the entire
network due to less contamination onto paths to other sinks.

%%%%%%%%%%%%%%%%%%%%%%%%%%%%%%%%%%%%%%%%%%%%%%%%%%%%%%%%%%%%%%%%%%%%%%%%%%%%%%%%
\section{An Algorithm for Assigning Streams}
\label{sec-streamassignalg}

We now give an algorithm for solving the non-uniform demand stream
assignment problem.  Again, we restrict ourselves to the case where
intermediate decoding does not occur within the network; that is, even
though nodes within the network may encode information, there is no
preliminary decoding (and removal of streams) except at the sink
nodes which are receiving information.  Equivalently, the output of
any node can only be more contaminated (but not less contaminated)
than any of the inputs to that node.  Admittedly, precluding the
removal of streams within the network does limit the solution space.
However, since the goal is to maximize the data rate to each sink,
solutions which do remove streams must be able to compensate for the
loss of data rate due to the removal of streams with some other
benefit (e.g., less contamination further down the network).

Our goal is to be able to assign streams to paths---with guarantees
that the contamination will be decodable---while maximizing the number
of streams transmitted.  We give a method which guarantees saturation
and decodability, by utilizing a polynomial transformation to graph
coloring.  We first give the transformation of the network graph into
a coloring graph in Algorithm~\ref{alg:color_graph_transform}, and
then give the algorithm for finding the saturating and decodable
solution in Algorithm~\ref{alg:graph_color_alg}.

\begin{algorithm}
\renewcommand{\captionfont}{\footnotesize}
\caption{Transformation of network graph into coloring graph}
\label{alg:color_graph_transform}
{\footnotesize
\begin{algorithmic}[1]
\REQUIRE Original network graph $G = (V,E)$, decomposed into
edge-disjoint paths $\mathcal{P}_j = \{p_{jk} \, | \,
k=1,\ldots,n_j\}$ for each sink~$j$
\STATE \label{it:vhat}
Create vertices $v_{jk} \in \hat{V}$, for $k=1,\ldots,n_j$, which are
associated with paths $p_{jk} \in \mathcal{P}_j$.  For each~$j$,
introduce $n-n_j$ additional vertices $w_{jk} \in \hat{V}$.  ($v_{jk}$
are \emph{regular vertices} and $w_{jk}$ are \emph{fictitious
vertices}.) Let
\[
\hat{V}_j = \{v_{jk} \, | \, 1 \leq k \leq n_j\} \cup \{w_{jk} \, | \,
            1 \leq k \leq n-n_j\}
\mbox{.}
\]
Call $\hat{V}_j$ the \emph{sink subgraph} associated with sink~$j$.
Then $\hat{V} = \bigcup_{j=1}^{t} \hat{V}_j$.
\STATE \label{it:ehat_complete}
For each~$j$, connect all vertices~$v_{jk}$ and vertices~$w_{jk}$
together into a clique.  Specifically, for each~$j$, let
\begin{eqnarray*}
\hat{E}_j^{\mathrm{complete}}
  & = & \{(v_{jk}, v_{jk'}) \, | \, 1 \leq k < k' \leq n_j\} \cup \\
  && \{(w_{jk}, w_{jk'}) \, | \, 1 \leq k < k' \leq n-n_j\} \cup \\
  && \{(v_{jk}, w_{jk'}) \, | \, 1 \leq k \leq n_j, 1 \leq k' \leq
     n-n_j\}
\mbox{.}
\end{eqnarray*}
\STATE \label{it:ehat_overlaps}
For each vertex~$v_{jk}$, connect~$v_{jk}$ to vertices~$w_{j'k'}$ for
all $k'=1,\ldots,n-n_{j'}$, if path~$p_{jk}$ contaminates onto some
path to sink~$j' \ne j$.  For each~$j$, call
\begin{eqnarray*}
\lefteqn{\hat{E}_j^{\mathrm{overlaps}} = } \\
&& \bigcup_{k=1}^{n_j} \{(v_{jk}, w_{j'k'}) \, | \, 1 \leq k' \leq
   n-n_{j'} \ \mathrm{if} \ \exists \tilde{k} \ \mathrm{s.t.} \ p_{j'
   \tilde{k}} \in \mathcal{D}_{jk}\}
\mbox{.}
\end{eqnarray*}
That is, if path~$p_{jk}$ to sink~$j$ contaminates onto some
path~$p_{j' \tilde{k}}$ to sink~$j'$, then node~$v_{jk}$ must connect
to all $n-n_{j'}$ fictitious vertices associated with sink~$j'$.
\STATE \label{it:ehat}
Let $\hat{E} = \bigcup_{j=1}^{t} (\hat{E}_j^{\mathrm{complete}} \cup
\hat{E}_j^{\mathrm{overlaps}})$.
\RETURN coloring graph $\hat{G} = (\hat{V}, \hat{E})$
\end{algorithmic}
} % end \footnotesize
\end{algorithm}

\begin{algorithm}
\renewcommand{\captionfont}{\footnotesize}
\caption{Non-uniform demand stream assignment}
\label{alg:graph_color_alg}
{\footnotesize
\begin{algorithmic}[1]
\REQUIRE Directed acyclic network graph $G = (V,E)$, source~$s$, and
sinks $j \in \{1,2,\ldots,t\}$
\STATE \label{it:edge_disjoint}
For each sink~$j$, find a set of edge-disjoint paths from~$s$ to~$j$.
Call the set of these paths $\mathcal{P}_j = \{p_{jk} \, | \,
k=1,\ldots,n_j\}$, where there are~$n_j$ such paths.  Let $n = \max_j
n_j$.
\STATE \label{it:create_graph}
Using Algorithm~\ref{alg:color_graph_transform}, construct coloring
graph $\hat{G}$ from $\mathcal{P} = \bigcup_{j=1}^{t} \mathcal{P}_j$.
\STATE \label{it:color}
Color~$\hat{G}$ using exactly~$n$ colors.  Let~$c_{jk}$ be the color
of~$v_{jk}$ in~$\hat{G}$.
\STATE \label{it:assign_streams}
For each path $p_{jk} \in \mathcal{P}$, assign stream~$c_{jk}$ to that
path.  (Each path $p_{jk} \in \mathcal{P}$ in the network graph~$G$ is
assigned the stream given by the color of its associated vertex
$v_{jk} \in \hat{V}$ in the coloring graph~$\hat{G}$.)
\STATE \label{it:netcode}
In the network graph, at each node where the inputs to the node are
different streams, send as the output of the node a combination of the
data from the input streams (e.g., linear combination, or some other
combining method)---taking care that no input streams are nullified in
the node output.
\end{algorithmic}
} % end \footnotesize
\end{algorithm}

The coloring graph~$\hat{G}$ can be interpreted with respect to the
solution found by Algorithm~\ref{alg:graph_color_alg}.  The
vertices~$v_{jk}$, $k=1,\ldots,n_j$, correspond to paths in the
original network, and so are called regular vertices.  The additional
vertices~$w_{jk}$, $k=1,\ldots,n-n_j$, correspond to fictitious paths,
indicating the streams which are \emph{not} assigned to paths leading
to sink~$j$---hence the name fictitious vertices.  The edges
in~$\hat{E}_j^\mathrm{complete}$ form a complete subgraph among
all~$n_j$ regular vertices associated with sink~$j$, guaranteeing
saturation.  (In fact, the entire induced subgraph of $\hat{V}_j$ is a
clique.) For~$\hat{E}_j^\mathrm{overlaps}$, the edges $(v_{jk},
w_{j'k'})$ connect the vertices from sink~$j$ to the vertices of
sink~$j'$ if there is some overlap on paths toward these two sinks;
these edges force a relationship between the streams assigned on paths
to one sink and streams assigned on paths to other sinks, providing
decodability.

Because $n = \max_j n_j$, there must exist at least one clique of
size~$n$ (associated with the induced subgraph of
$\hat{V}_{j^{\star}}$, where $j^{\star} = \arg\max_j n_j$).  Thus,
$\hat{G}$~can not be colored with fewer than~$n$ colors, so
step~\ref{it:color} of Algorithm~\ref{alg:graph_color_alg} is
equivalent to coloring~$\hat{G}$ with \emph{at most}~$n$ colors.  If
the minimum coloring solution of~$\hat{G}$ requires more than~$n$
colors, then the following lemma tells us that decodability has been
violated.

\begin{lemma}
\label{lem:coloring_violation}
In the equivalent coloring graph constructed from
Algorithm~\ref{alg:color_graph_transform}, if the chromatic number
$\chi(\hat{G}) > n$, then the original network is not decodable.

\begin{proof}
If $\chi(\hat{G}) > n$ and every vertex is a member of at least one
induced clique of size~$n$, then there must exist two cliques of
size~$n$ such that there is at least one edge connecting these two
cliques.  Moreover, because the coloring graph must be colored with
more than~$n$ colors, some pair $(j,j')$ of connected cliques (each
clique associated with a different sink) must satisfy the following
condition:  If clique~$j$ does not have color~$c$ within it, then its
fictitious vertices must be connected to a regular vertex in
clique~$j'$ which has color~$c$.  In this case, sink~$j$ can not
decode color~$c$ even though some path to~$j$ has contamination~$c$
from a path to~$j'$, and so decodability is violated.
\end{proof}
\end{lemma}

\begin{theorem}
\label{thm:sat_decod_cond}
Algorithm~\ref{alg:graph_color_alg} succeeds in coloring the
equivalent coloring graph with exactly $n$ colors if and only if the
original network graph has a decodable and saturating solution with no
intermediate decoding.

\begin{proof}
It is clear that $\chi(\hat{G}) = n$ is necessary for the solution to
be decodable and saturating.  From Lemma~\ref{lem:coloring_violation},
we know that if the network is decodable, then the equivalent coloring
graph must have $\chi(\hat{G}) \leq n$.  Now consider
sink~$j^{\star}$, where $j^{\star} = \arg\max_j n_j$.  If the network
is saturating, then sink~$j^{\star}$ must be able to receive $n =
n_{j^{\star}}$ distinct streams.  That is, the clique associated with
sink~$j^{\star}$ must be colored with at least $n$~colors.  This gives
us $\chi(\hat{G}) \geq n$.  Thus, $\chi(\hat{G}) = n$.

Next we prove sufficiency of $\chi(\hat{G}) = n$ for a decodable and
saturating solution.  For the coloring to be valid, if path~$p_{jk}$
contaminates onto \emph{any} path~$p_{j' \tilde{k}}$ to sink~$j'$,
then by construction of~$\hat{E}^\mathrm{overlaps}$, none of the
fictitious vertices~$w_{j'k'}$ associated with sink~$j'$ may have the
same color as vertex~$v_{jk}$ (call this color~$c_{jk}$).  Because
$\chi(\hat{G}) = n$ and each sink subgraph is an induced clique of
size~$n$, some regular vertex~$v_{j'k'}$ to sink~$j'$ must be
colored~$c_{jk}$.  Equivalently, contamination due to path~$p_{jk}$
onto path~$p_{j' \tilde{k}}$ has been assigned to some path~$p_{j'k'}$
to sink~$j'$ (and this is true for all possible contaminations), so by
Theorem~\ref{thm:sat_decode_cond} the solution is decodable.
Moreover, if $\chi(\hat{G}) = n$, then by construction
of~$\hat{E}^\mathrm{complete}$, all paths have assigned streams, and
the assigned streams are distinct for different paths to the same
sink.  Thus, the solution is saturating.
\end{proof} 
\end{theorem}

Theorem~\ref{thm:sat_decod_cond} gives necessary and sufficient
conditions for a saturating and decodable solution to exist.
Moreover, the stream assignment algorithm tells us how to allocate
streams in order to construct this solution.

\begin{example}[Extended Butterfly Network]
\label{ex:extended_butterfly_network}
Figure~\ref{fig:extended_butterfly_example} shows the result of
Algorithm~\ref{alg:graph_color_alg} on the extended butterfly network.
\begin{figure}[htbp]
\centering
\subfigure[Path decomposition for extended butterfly network]{
\psfrag{s}[c][c]{\footnotesize $s$}
\psfrag{u}[c][c]{\footnotesize $u$}
\psfrag{v}[c][c]{\footnotesize $v$}
\psfrag{w}[c][c]{\footnotesize $w$}
\psfrag{x}[c][c]{\footnotesize $x$}
\psfrag{y}[c][c]{\footnotesize $y$}
\psfrag{z}[c][c]{\footnotesize $z$}
\psfrag{t1}[c][c]{\footnotesize $t_{\mathscriptsize{1}}$}
\psfrag{t2}[c][c]{\footnotesize $t_{\mathscriptsize{2}}$}
\psfrag{p11}[c][c]{\footnotesize $p_{\mathscriptsize{1,1}}$}
\psfrag{p12}[c][c]{\footnotesize $p_{\mathscriptsize{1,2}}$}
\psfrag{p21}[c][c]{\footnotesize $p_{\mathscriptsize{2,1}}$}
\psfrag{p22}[c][c]{\footnotesize $p_{\mathscriptsize{2,2}}$}
\psfrag{p23}[c][c]{\footnotesize $p_{\mathscriptsize{2,3}}$}
\includegraphics[height=1.3in]{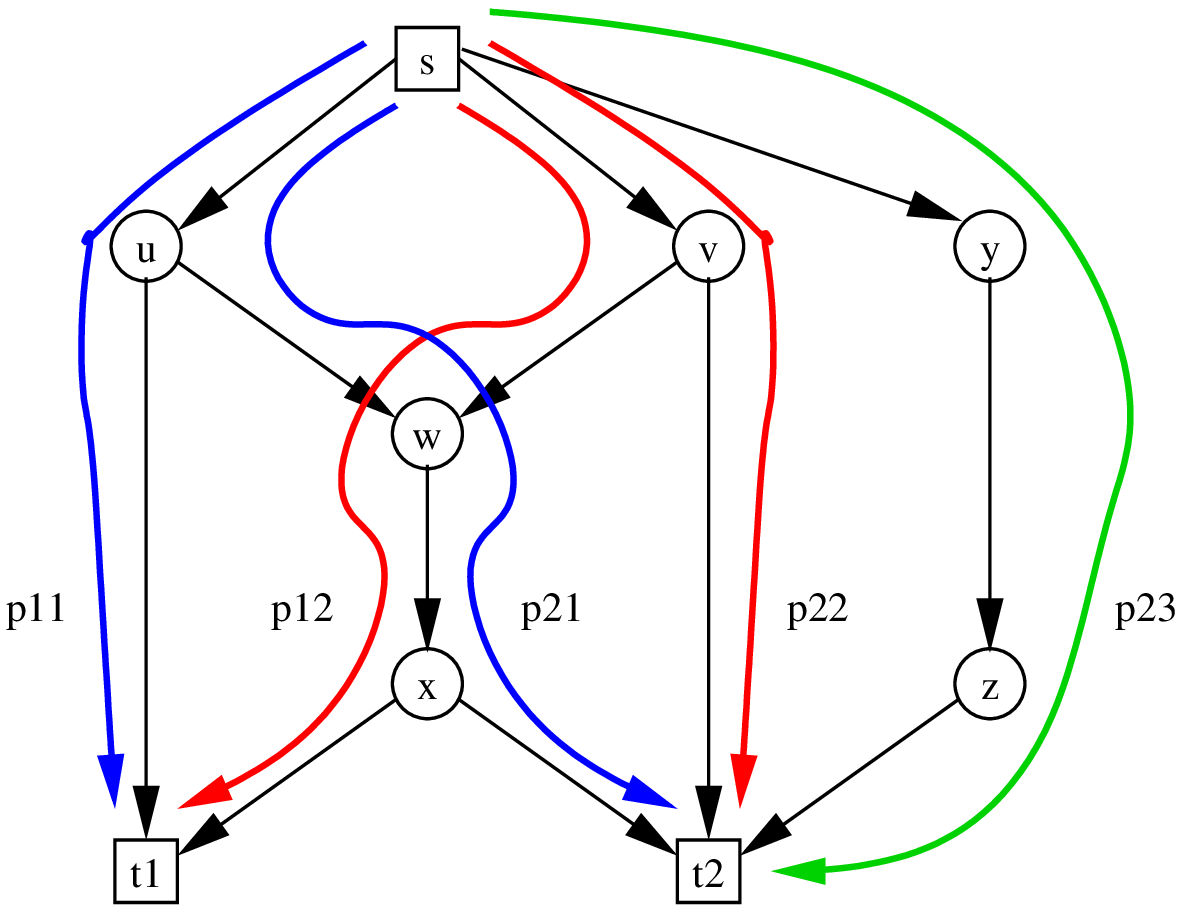}
\label{fig:extended_butterfly_paths}
}
\hspace{0.1in}
\subfigure[Equivalent coloring graph for extended butterfly network]{
\psfrag{v11}[c][c]{\footnotesize $v_{\mathscriptsize{1,1}}$}
\psfrag{v12}[c][c]{\footnotesize $v_{\mathscriptsize{1,2}}$}
\psfrag{w11}[c][c]{\footnotesize $w_{\mathscriptsize{1,1}}$}
\psfrag{v21}[c][c]{\footnotesize $v_{\mathscriptsize{2,1}}$}
\psfrag{v22}[c][c]{\footnotesize $v_{\mathscriptsize{2,2}}$}
\psfrag{v23}[c][c]{\footnotesize $v_{\mathscriptsize{2,3}}$}
\includegraphics[height=0.75in]{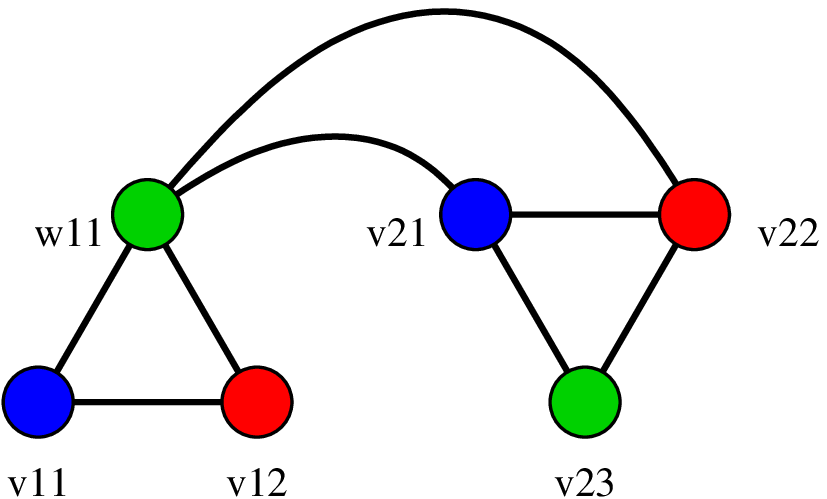}
\label{fig:extended_butterfly_coloring_graph}
}
\renewcommand{\captionfont}{\footnotesize}
\caption{The path decomposition for the extended butterfly network is
shown in Figure~\ref{fig:extended_butterfly_paths}.  The color of each
path indicates the stream which should be assigned to that path.
Figure~\ref{fig:extended_butterfly_coloring_graph} shows the
equivalent coloring graph for the extended butterfly network.
The graph can be colored with~$3$ colors, so it gives a saturating and
decodable solution to the original network.  The colors correspond to
the streams which should be assigned to the paths.
}
\label{fig:extended_butterfly_example}
\end{figure}
\end{example}

\subsection{Shortcomings}
\label{subsec-alg_shortcomings}

Assuming the correct set of edge-disjoint paths are chosen in the
first step of the algorithm, then if the solution exists it will be
found.  However, we do not address the proper selection of
edge-disjoint paths, even though there may be multiple path
decompositions, where some decompositions lead to sub-optimal
assignments.  For example, it is possible to construct a
counterexample network with a given path decomposition, where
switching a single edge for one path greatly increases allowed
throughput.

The algorithm determines---for a given set of paths---whether or
not~$n$ streams can be assigned, but to determine if some $\bar{n} >
n$ streams can be assigned, additional fictitious vertices need to be
introduced.  For each sink, an additional $\bar{n} - n$ fictitious
vertices must be introduced in order to get sink subgraphs of size
$\bar{n}$.  The relationship between these larger graphs and the
original coloring graph is unknown, and it is possible that no matter
how large $\bar{n}$ is chosen, it will still be impossible to find a
saturating and decodable assignment with $\bar{n}$ streams.  If one
wishes to determine how many additional available streams will
guarantee saturation and decodability, $\bar{n}$ could be increased
without bound while searching for a possible stream assignment.

In fact, it is possible to have networks where there does not exist a
saturating solution which is also decodable, no matter how large the
available stream set is.  This can occur when there is too much
overlap but not enough available paths to remove the contamination.
For example, consider a two sink case, where $n_1 = 1$ and $n_2 = 2$.
Suppose both paths of sink~$2$ overlap the path to sink~$1$ at some
link[s].  No matter what the stream assignment (or how large the space
of possible streams) for sink~$1$, it will never be able to decode out
both contaminants if saturation for both sinks is required.

Thus, our algorithm only works for the restrictive case where no
intermediate decoding is allowed, yet all paths to sinks must be
saturating.  Either loosening the saturation or the no intermediate
decoding restrictions would be beneficial, but at the moment, the
algorithm relies on both conditions.

Another issue to keep in mind is that because coloring is an
NP-complete problem, there are no known polynomial time algorithms
which will perform the coloring step (unless $P = NP$).  Additionally,
there are no good approximation algorithms known for the graph
coloring problem (see~\cite{karger:approx_graph_coloring_sdp} for
algorithms which can color $n$-colorable graphs with number of colors
logarithmic in the number of vertices of the graph, but with no
guarantees based on the actual chromatic number $n$).  Even if there
were good approximation algorithms, an approximation algorithm might
not be enough to answer the question of whether or not a saturating
solution exists (i.e., whether or not $\chi(\hat{G}) = n$) since we
require finding the chromatic number exactly.  One might conjecture
that because the coloring graph~$\hat{G}$ is carefully constructed, it
might have some special structure which would allow for a
polynomial-time coloring algorithm.  In the next section, we give some
structural properties of the coloring graph which can lead to a
polynomial-time coloring, but we also show a counterexample network
where this particular structural analysis is not sufficient to prove
polynomial-time solvability.

%%%%%%%%%%%%%%%%%%%%%%%%%%%%%%%%%%%%%%%%%%%%%%%%%%%%%%%%%%%%%%%%%%%%%%%%%%%%%%%%
\section{Efficiently-Solvable Non-Uniform Demand Problems}
\label{sec-effnonunifdemand}

The coloring step in the stream assignment problem is problematic, as
the graph coloring problem is NP-complete and so in general no known
polynomial-time algorithm can solve the problem.  However, if we
restrict our class of demand problems to only those for which the
corresponding coloring graph is polynomial-time solvable, then such
demand problems will also be polynomial-time solvable.  Specifically,
we consider the class of graphs known as Berge graphs, which are
graphs characterized by the absences of both odd holes (induced cycles
of odd length at least 5) and odd antiholes (complements of odd
holes).  Then by the strong perfect graph
theorem~\cite{chudnovsky:strong_perfect_graph_theorem}, a Berge graph
is also a perfect graph, so the chromatic number of a Berge graph is
equal to the size of its maximum clique.

This fact is useful in finding solutions to the non-uniform demand
problem because in our formulation, the maximum clique is easily
found, so if the coloring graph is Berge, then the chromatic number is
also readily determined.  The main result of this section is that we
can find the maximum clique for the coloring graph of
Algorithm~\ref{alg:graph_color_alg} in polynomial time and hence also
its chromatic number if the coloring graph is Berge.  We first prove
some preliminary results.
\begin{lemma}
\label{lem:no_clique_three_sink_cliques}
Any induced clique consisting of vertices from different sink
subgraphs can only consist of vertices from at most two sink
subgraphs.  That is, it is impossible to induce a complete subgraph
consisting of at least one vertex from each of $\hat{V}_j$,
$\hat{V}_{j'}$, and $\hat{V}_{j''}$.

\begin{proof}
For vertices belonging to different sink subgraphs, i.e., $v \in
\hat{V}_j$ and $v' \in \hat{V}_{j'}$ where $j \ne j'$, either~$v$ is
regular and~$v'$ is fictitious, or~$v$ is fictitious and~$v'$ is
regular.  Regular vertices are not connected to regular vertices, nor
are fictitious vertices connected to fictitious vertices---unless they
belong to the same sink subgraph.  Any complete subgraph consisting of
at least one vertex from each of $\hat{V}_j$, $\hat{V}_{j'}$, and
$\hat{V}_{j''}$ must contain at least two vertices of the same type
from different sink subgraph (e.g., two regular vertices and one
fictitious vertex, where each vertex is from a different sink
subgraph).  However, such a scenario can not exist, as that implies
that two vertices of the same type but from different sink subgraphs
are connected.
\end{proof}
\end{lemma}

The preceding lemma tells us that in order to find the maximum clique
in the coloring graph, all we need to do is search for induced cliques
pairwise between sink subgraphs.  We can select sink subgraphs two at
a time and determine the largest induced complete subgraph consisting
only of vertices from these two sink subgraphs.  This procedure
requires solving~$\binom{t}{2}$ subproblems, where each subproblem can
be performed in time which is polynomial in~$n$.
\begin{lemma}
\label{lem:max_clique_pairwise_polytime}
For a pair of sink subgraphs~$\hat{V}_j$ and~$\hat{V}_{j'}$, finding
the maximum induced complete subgraph of $\hat{V}_j \cup \hat{V}_{j'}$
takes time polynomial in the sink subgraph size~$n$.

\begin{proof}
If sinks~$j$ and~$j'$ do not have any overlapping paths, then
because~$\hat{V}_j$ and~$\hat{V}_{j'}$ are disjoint, the maximum
induced complete subgraph of $\hat{V}_j \cup \hat{V}_{j'}$
is~$\hat{V}_j$ (or~$\hat{V}_{j'}$), which has size~$n$.
Disjointness of~$\hat{V}_j$ and~$\hat{V}_{j'}$ can be checked by
considering each vertex~$v_{jk} \in \hat{V}_j$ and seeing if it has an
edge to any vertex in $\hat{V}_{j'}$.  This requires $n$ steps.

If a path~$p_{jk}$ to sink~$j$ contaminates onto some path to
sink~$j'$, then the regular vertex~$v_{jk}$ is connected to all of the
fictitious vertices of~$\hat{V}_{j'}$.  Thus, the largest induced
complete subgraph consisting of both regular vertices from~$\hat{V}_j$
and fictitious vertices from~$\hat{V}_{j'}$ has size $m_{j,j'} +
(n-n_{j'})$, where~$m_{j,j'}$ is the number of regular vertices
of~$\hat{V}_j$ which are connected to the fictitious vertices
of~$\hat{V}_{j'}$.  (Recall that $n-n_{j'}$ is the number of
fictitious vertices of~$\hat{V}_{j'}$.)  Computing~$m_{j,j'}$
takes~$O(n)$ time, as it merely requires counting up the number of
regular vertices of~$\hat{V}_j$ that are connected to the fictitious
vertices of~$\hat{V}_{j'}$.  Equivalently, $m_{j,j'}$ can be computed
by counting the number of paths to sink~$j$ which contaminate onto
some path to sink~$j'$.  Of course, the largest induced complete
subgraph of $\hat{V}_j \cup \hat{V}_{j'}$ may instead consist of
regular vertices from~$\hat{V}_{j'}$ and fictitious vertices
from~$\hat{V}_j$; by a similar argument, finding such a subgraph also
takes polynomial time.  Then we can find the maximum induced complete
subgraph of $\hat{V}_j \cup \hat{V}_{j'}$, and the size of this induced
subgraph is $n + \max(0, m_{j,j'} - n_{j'}, m_{j',j} - n_j)$.
\end{proof}
\end{lemma}

From this, it can be readily shown that finding the maximum clique in
the coloring graph is polynomial-time.  We can then conclude that it
is possible to determine the existence of a saturating and decodable
solution (again, disregarding intermediate decoding) in polynomial
time.
\begin{theorem}
\label{thm:berge_polytime}
For a particular non-uniform demand scenario, if the associated
coloring graph is a Berge graph, then it is a polynomial-time
operation to determine whether or not there exists a saturating and
decodable solution which does not require intermediate decoding.
Moreover, if the solution exists, it can be found using the
non-uniform demand stream assignment algorithm
(Algorithm~\ref{alg:graph_color_alg}).

\begin{proof}
From Lemma~\ref{lem:max_clique_pairwise_polytime}, we know that
finding the maximum induced clique between two sink subgraphs is a
polynomial time operation.  Thus, finding the maximum clique of the
coloring graph takes polynomial time, as it consists of solving
$\binom{t}{2} = \frac{t(t-1)}{2}$ such subproblems.  Because this
coloring graph is a Berge graph, then its chromatic number can be
found in polynomial time, since the chromatic number is equal to the
maximum clique size.  From Theorem~\ref{thm:sat_decod_cond}, we can
then determine if the original network graph has a saturating and
decodable solution.  Not only that, but if the coloring requires no
more than~$n$ colors, then the coloring found from the stream
assignment algorithm immediately gives the non-uniform demand
solution.
\end{proof}
\end{theorem}

The interpretation of Theorem~\ref{thm:berge_polytime} is that for
coloring graphs which are Berge graphs, if we find that the maximum
clique has size~$n$, then we can conclude that the coloring graph can
be colored with~$n$ colors, and so the original non-uniform demand
network coding problem has a saturating and decodable solution.  If,
however, the maximum clique has size greater than~$n$, then we can
also conclude that no saturating and decodable solution exists---at
least no solution which does not require intermediate decoding.  This
result is particularly promising, as we can then quickly enumerate a
sufficient condition under which a saturating and decodable solution
can be found in polynomial time---specifically, if the associated
coloring graph is Berge.

One might ask if the step of determining whether or not a graph is
Berge might be a difficult task, as any difficulties in doing so would
outweigh any benefits gained by solving the demand problem
efficiently.  However, a polynomial-time algorithm for recognizing
Berge graphs does exist~\cite{chudnovsky:recog_berge}.  Consequently,
if it is determined that the coloring graph is a Berge graph, then the
stream assignment algorithm can be used to find the saturating and
decodable solution to the non-uniform demand problem in polynomial
time.  Or, if it is determined that the coloring graph is not a Berge
graph, then some other, perhaps superpolynomial time, algorithm will
be needed to perform the coloring step.

However, non-uniform demand scenarios which lead to non-Berge coloring
graphs do exist.  We give an example.
\begin{example}[Network with non-Berge coloring graph]
\label{ex:non_berge_coloring_graph}
Consider the network given in Figure~\ref{fig:non_berge_example}.  Its
corresponding coloring graph has an odd hole of length~$5$, induced by
the vertices $v_{1,1}$, $w_{1,1}$, $v_{2,1}$, $v_{2,2}$, and
$w_{3,1}$, so it is not Berge.  However, a valid coloring of size
$n=3$ does exist.
\begin{figure}[htbp]
\centering
\subfigure[Network with non-Berge coloring graph]{
\psfrag{s}[c][c]{\small $s$}
\psfrag{x11}[c][c]{\tiny $x_{\mathsupertiny{11}}$}
\psfrag{x12}[c][c]{\tiny $x_{\mathsupertiny{12}}$}
\psfrag{x21}[c][c]{\tiny $x_{\mathsupertiny{21}}$}
\psfrag{x22}[c][c]{\tiny $x_{\mathsupertiny{22}}$}
\psfrag{x31}[c][c]{\tiny $x_{\mathsupertiny{31}}$}
\psfrag{x32}[c][c]{\tiny $x_{\mathsupertiny{32}}$}
\psfrag{x41}[c][c]{\tiny $x_{\mathsupertiny{41}}$}
\psfrag{x42}[c][c]{\tiny $x_{\mathsupertiny{42}}$}
\psfrag{x43}[c][c]{\tiny $x_{\mathsupertiny{43}}$}
\psfrag{t1}[c][c]{\footnotesize $t_{\mathscriptsize{1}}$}
\psfrag{t2}[c][c]{\footnotesize $t_{\mathscriptsize{2}}$}
\psfrag{t3}[c][c]{\footnotesize $t_{\mathscriptsize{3}}$}
\psfrag{t4}[c][c]{\footnotesize $t_{\mathscriptsize{4}}$}
\includegraphics[height=1.7in]{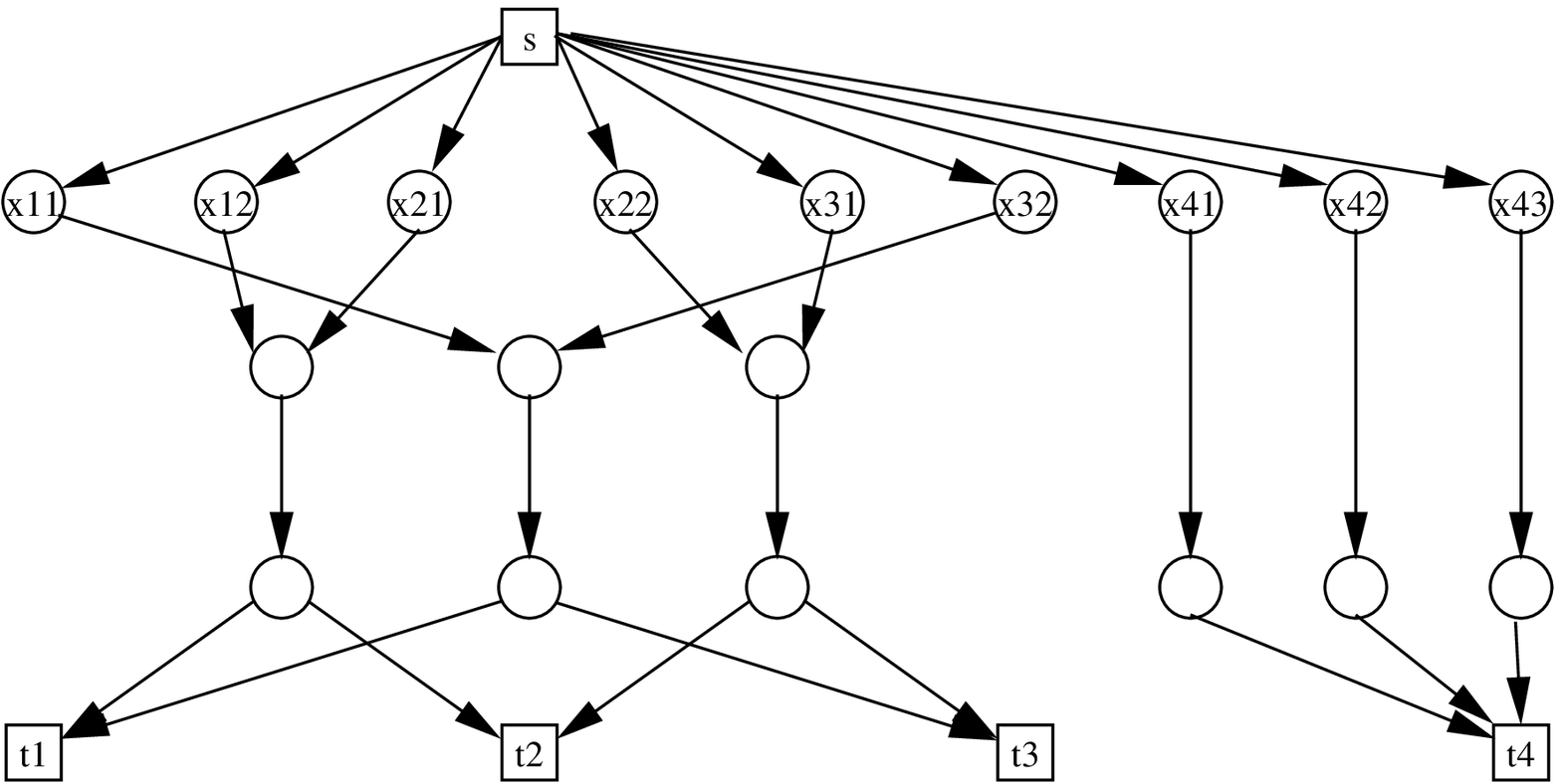}
\label{fig:non_berge_network}
}
\\
\subfigure[Coloring graph for non-Berge network]{
\psfrag{v11}[c][c]{\footnotesize $v_{\mathscriptsize{1,1}}$}
\psfrag{v12}[c][c]{\footnotesize $v_{\mathscriptsize{1,2}}$}
\psfrag{w11}[c][c]{\footnotesize $w_{\mathscriptsize{1,1}}$}
\psfrag{v21}[c][c]{\footnotesize $v_{\mathscriptsize{2,1}}$}
\psfrag{v22}[c][c]{\footnotesize $v_{\mathscriptsize{2,2}}$}
\psfrag{w21}[c][c]{\footnotesize $w_{\mathscriptsize{2,1}}$}
\psfrag{v31}[c][c]{\footnotesize $v_{\mathscriptsize{3,1}}$}
\psfrag{v32}[c][c]{\footnotesize $v_{\mathscriptsize{3,2}}$}
\psfrag{w31}[c][c]{\footnotesize $w_{\mathscriptsize{3,1}}$}
\psfrag{v41}[c][c]{\footnotesize $v_{\mathscriptsize{4,1}}$}
\psfrag{v42}[c][c]{\footnotesize $v_{\mathscriptsize{4,2}}$}
\psfrag{v43}[c][c]{\footnotesize $v_{\mathscriptsize{4,3}}$}
\includegraphics[height=0.75in]{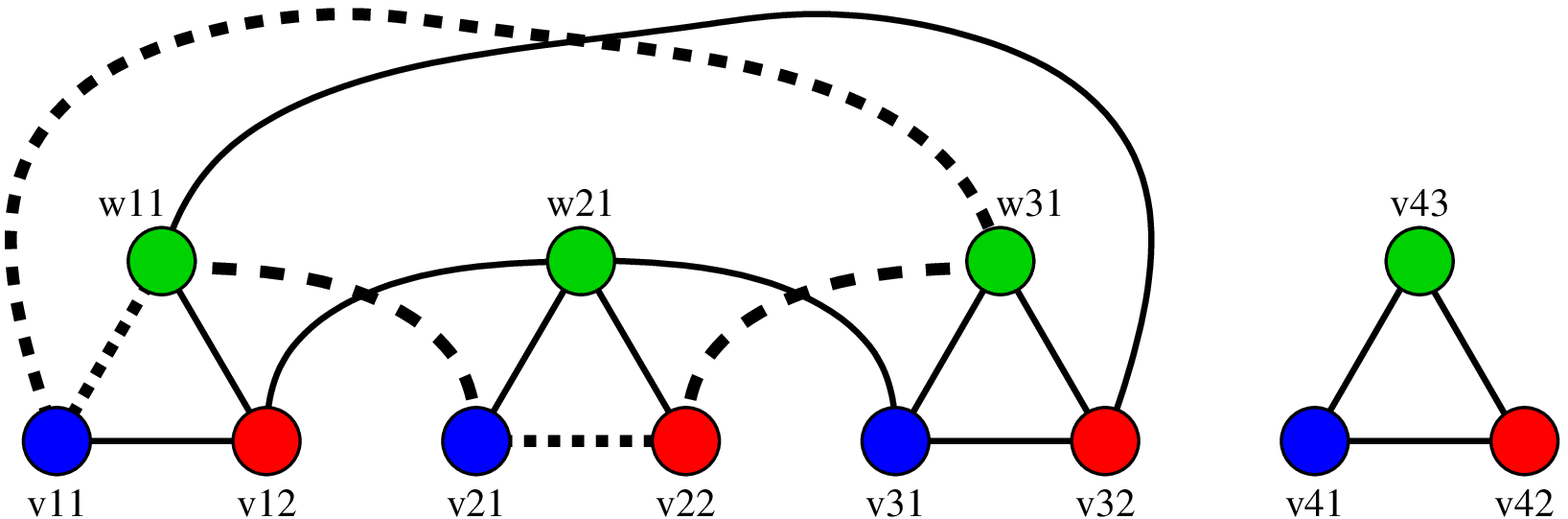}
\label{fig:non_berge_network_coloring_graph}
}
\renewcommand{\captionfont}{\footnotesize}
\caption{Figure~\ref{fig:non_berge_network} shows an example of a
network whose coloring graph is not Berge.  For each~$j$ and~$k$,
path~$p_{j,k}$ is the path which passes through node~$x_{jk}$ on its
way to sink~$t_j$.  Figure~\ref{fig:non_berge_network_coloring_graph}
shows the equivalent coloring graph for this path decomposition.  The
dotted lines indicate the induced subgraph of vertices $v_{1,1}$,
$w_{1,1}$, $v_{2,1}$, $v_{2,2}$, and $ w_{3,1}$; the induced subgraph
is an odd hole of length~$5$, so this coloring graph is not a Berge
graph.  It should be noted, however, that a coloring using $n=3$
colors does exist for this graph, and is as shown.}
\label{fig:non_berge_example}
\end{figure}
\end{example}

%%%%%%%%%%%%%%%%%%%%%%%%%%%%%%%%%%%%%%%%%%%%%%%%%%%%%%%%%%%%%%%%%%%%%%%%%%%%%%%%
\section{Conclusion}
\label{sec-conclusion}

In this paper, we have considered the non-uniform demand network
coding problem, where the sinks are allowed to receive data at unequal
rates.  We give an algorithm for finding network coding solutions
which satisfy the decodability and saturation properties.
Additionally, we show that for certain types of networks, i.e., those
which can be transformed into equivalent Berge graphs, our algorithm
can find the solution in polynomial time.  Moreover, it will be
difficult to do much better for the general case, as the non-uniform
demand problem is NP-hard, even when demands are restricted to only
those which are saturating.

Our results can be interpreted relative to the case of equal-rate
multicast, where network coding solutions can always be found in
polynomial time~\cite{jaggi:polytime_algs_multicast_code_constr}.  Our
algorithm relies on introducing fictitious vertices in the coloring
graph, which corresponds to introducing fictitious paths in the
original network.  These fictitious paths do not overlap anywhere with
any other paths, but only serve to bring the total number of paths to
each sink up to~$n$.  We can perform multicast on the expanded network
consisting of the original network plus the network induced by the
fictitious paths; in this expanded network, each sink is guaranteed to
receive~$n$ streams of information.  The main contribution of our work
is that using our algorithm, we can directly specify that the subset
of paths in the expanded network corresponding to paths in the
original network is assigned mutually-decodable streams, without
needing any of the information transmitted on the fictitious paths.
From the perspective of the equal-rate multicast problem on the
expanded graph, the interpretation of our algorithm is that it
provides a partitioning of information between data transmitted on the
original network paths and data transmitted on the fictitious paths.

There are a few issues which require further study.  We have not
considered the optimal selection of paths; our assumption is that the
set of paths we use are the ones which give the demand solution if it
exists.  The optimal selection of paths is a challenging problem in
itself, as it requires knowing what types of demand solutions may
arise from the particular choice of paths.  When implementing this
algorithm, heuristics---such as minimizing the number of overlapping
links on paths to different sinks---will most likely be sufficient.
We also mention that although our conditions guarantee that a network
code will exist if our algorithm finds a solution, the actual
construction of the network code is not detailed.  In the case of no
intermediate decoding, linear codes will be sufficient.  However, if
intermediate decoding were to be allowed, then we must be more
careful, as it has been shown that sometimes nonlinear codes are
required to solve certain other network coding
problems~\cite{dougherty:insuff_linear_coding}.

Our approach considers network coding scenarios which are scalar,
where the same code is employed during every time period.  Although
this allows for a wide variety of codes and is also practically
implementable, there are certain network coding problems where vector
solutions (i.e., solutions where the network code may be different at
each time period) exist, but scalar solutions do
not~\cite{medard:on_coding_non_multicast}.  One avenue of inquiry
would be the adaptation of our algorithms to find vector solutions in
the cases where scalar solutions do not exist; this should be possible
by augmenting our network graphs to also include a time dimension.
However, characterizing the set of networks with
polynomial-time-solvable vector solutions (but no scalar solutions)
will require more work.

%%%%%%%%%%%%%%%%%%%%%%%%%%%%%%%%%%%%%%%%%%%%%%%%%%%%%%%%%%%%%%%%%%%%%%%%%%%%%%%%
\appendix %[NP Completeness]
\label{app-npcomp}

We give an alternate proof of NP-completeness of the non-uniform
demand network coding problem, via a polynomial reduction from a
general graph coloring problem.
Unlike~\cite{cassuto:non_uniform_demands}, in which the network demand
problems shown to be NP-hard  do not have fully saturated demands, our
proof considers sink demands in which saturation must occur.  The
coloring problem which we consider is the following:  Given an
undirected graph $\hat{G} = (\hat{V}, \hat{E})$, is there a coloring
using $n$ [or fewer] colors?  We first give the reduction, followed by
a proof that the reduction leads to an equivalent problem.

\begin{algorithm}
\renewcommand{\captionfont}{\footnotesize}
\caption{A reduction from general graph coloring to non-uniform demand
stream assignment}
\label{alg:reduc_from_graph_color}
{\footnotesize
\begin{algorithmic}[1]
\REQUIRE Undirected graph $\hat{G} = (\hat{V},\hat{E})$ to be colored
\STATE \label{it:constr_paths}
For each edge $e_j = (v_j,w_j) \in \hat{E}$, construct a sink~$j$ in
the network graph consisting of two paths~$p_{j,1}$ and~$p_{j,2}$.  If
two edges $e_j = (v,w_j) \in \hat{E}$ and $e_i = (v,w_i) \in \hat{E}$
share a vertex~$v$, then force the paths~$p_{j,1}$ and~$p_{i,1}$ to
overlap at some link.  Call the overlapping link in the network graph
by~$v$.  (If the shared vertex is~$w$ such that $e_j = (v_j,w) \in
\hat{E}$ and $e_i = (v_i,w) \in \hat{E}$, then force paths $p_{j,2}$
and $p_{i,2}$ to overlap at some link~$w$.)  Thus, vertices in the
coloring graph determine the intersections of paths in the network
graph---where the link of intersection occurs according to the vertex
in the coloring graph.
\STATE \label{it:same_paths}
Introduce another~$|V|$ sinks, with only a single path to each sink.
Label these sinks $1,2,\ldots,|V|$.  For a particular sink~$v$, call
the single path~$p_v$, and make path~$p_v$ intersect with all other
paths which cross through link~$v$ in the network graph.
\STATE \label{it:force_n_colors}
Introduce one additional sink, with~$n$ paths.  These~$n$ paths do not
intersect any paths defined in prior steps.
\STATE
Solve the non-uniform demand stream assignment problem on the
resulting network graph.
\end{algorithmic}
} % end \footnotesize
\end{algorithm}

Step~\ref{it:constr_paths} of the above algorithm sets up most of the
network graph.  Overlaps between paths reflect the fact that a vertex
can not be colored two different colors.  The addition of~$|V|$ sinks
in step~\ref{it:same_paths} forces the stream assignment algorithm to
assign the same color to all paths crossing through the same link~$v$;
otherwise, in the sinks with two paths, it may be possible that the
corresponding stream will be assigned to the path in the pair which
does not overlap at the considered link. Furthermore, the single sink
with~$n$ paths in step~\ref{it:force_n_colors} guarantees that at
least~$n$ different streams will be assigned.

\begin{lemma}
\label{lem:nudsa_equiv_coloring}
Performing non-uniform demand stream assignment on the network
digraph~$G$ formed from Algorithm~\ref{alg:reduc_from_graph_color} is
equivalent to coloring the original undirected graph~$\hat{G}$.

\begin{proof}
First we show that if there exists a coloring solution for~$\hat{G}$
using~$n$ colors, then there will also be a non-uniform demand
stream assignment on the constructed network graph with~$n$ streams.
To do so, start with a coloring solution.  For a particular
vertex~$v$ in the coloring graph, assign the stream corresponding to
the color of vertex~$v$ to all paths which intersect at the associated
link~$v$ in the network graph.  Because no path has more than one link
which has overlap, then there is no ambiguity about the stream which
is assigned to that path. Because graph coloring guarantees that the
vertices connected by an edge will be colored different colors, each
sink from step~\ref{it:constr_paths} will receive two paths that have
different stream assignments.  Thus, the solution is saturating.  Not
only that, in the network graph, any intersecting paths only intersect
at one link, so contamination is mitigated by assigning the same
stream to all paths which intersect at the same edge.  Thus, the
solution to the graph coloring with~$n$ colors gives a non-uniform
demand stream assignment for the constructed network graph using
exactly~$n$ distinct streams.  This stream assignment is both
saturating and decodable.

Conversely, assume that there exists a saturating and decodable stream
assignment to the non-uniform demand stream assignment problem on the
constructed network graph, which uses exactly~$n$ streams.  Then this
solution can be used to determine a graph coloring of the original
graph, with~$n$ colors.  To prove this, first consider the sinks which
have single paths.  From these sinks, suppose path~$p_v$ to sink~$v$
(associated with vertex~$v$ in the coloring graph) is assigned
stream~$c$.  By decodability, any other paths which intersect
path~$p_v$ must be carrying stream~$c$.  That is, for a path~$p'_j$
which intersects~$p_v$, then its pair path~$p''_j$ (i.e., toward the
same sink) is not the path of the pair which is carrying stream~$c$.
Otherwise, sink~$v$ would need to decode out the stream on path~$p'_j$
(which would be some $c' \ne c$), but sink~$v$ can not, since it is
only assigned to receive stream~$c$ from the single path~$p_v$.  Thus,
all paths associated with the same vertex~$v$ in the original coloring
graph must have the same stream assignment; this is the color assigned
to vertex~$v$.  Now, consider the sinks with paired paths.  Because
the non-uniform demand stream assignment solution is saturating, that
means that the two paths are assigned different streams.  This is
equivalent to the requirement that vertices which are connected by an
edge in the original coloring graph be assigned different colors.
Thus, the stream assignment on~$G$ using~$n$ streams gives a coloring
on~$\hat{G}$ using~$n$ colors.
\end{proof}
\end{lemma}

From the preceding construction, it is straightforward to determine
the complexity of the stream assignment problem.
\begin{theorem}
\label{thm:nudsa_np_complete}
The non-uniform demand stream assignment problem with saturated
demands is NP-complete.

\begin{proof}
First we show that the problem is in NP, by showing that it takes
polynomial time to check if a given solution to the non-uniform demand
stream assignment problem is feasible.  For every overlap link in the
network graph~$G$, check if the sinks receiving the paths crossing
through the link have additional streams assigned to them which are
the same as all the contaminations from that link.  This takes at most
$O(n|E|)$ time, where~$|E|$ is the number of links in the
corresponding network graph (and~$|E|$ can be as small as $4|\hat{E}|
+ 3|\hat{V}| + n$).

Next, we show that the transformation from the coloring graph to the
related network graph given in
Algorithm~\ref{alg:reduc_from_graph_color} is polynomial.  The network
graph has $|\hat{E}| + |\hat{V}| + 1$ sinks (more precisely, it has
$2|\hat{E}| + |\hat{V}| + n$ paths), so the reduction is polynomial.
Because any instance of graph coloring can be polynomially reduced to
a non-uniform demand stream assignment problem, and solutions can be
checked in polynomial time, the non-uniform demand stream assignment
problem is NP-complete.
\end{proof}
\end{theorem}

From the above arguments, we also conclude that non-uniform demand
stream assignment can be solved in polynomial time when $n = 2$ (i.e.,
when the maximum data rate to any sink is upper bounded by~$2$).  This
is because graph coloring is polynomial time (i.e., by searching for
bipartiteness in the graph) when only~$2$ colors are
allowed~\cite{garey:np_completeness}.

%%%%%%%%%%%%%%%%%%%%%%%%%%%%%%%%%%%%%%%%%%%%%%%%%%%%%%%%%%%%%%%%%%%%%%%%%%%%%%%%
\section*{Acknowledgements}
\label{sec-ack}

The authors would like to thank Amin Saberi for insightful
discussions, and William Wu and Caleb Lo for helpful comments which
improved the readability of this paper.  Part of this work was
completed while J. Koo was supported by a National Defense Science and
Engineering Graduate Fellowship.

%%%%%%%%%%%%%%%%%%%%%%%%%%%%%%%%%%%%%%%%%%%%%%%%%%%%%%%%%%%%%%%%%%%%%%%%%%%%%%%%

\bibliographystyle{IEEEtran} % use IEEEtran.bst style
\bibliography{IEEEabrv,references}

\end{document}